\documentclass[aoas,nameyear,dvips]{arximspdf}
\usepackage{dcolumn}
\usepackage{graphicx}

\doi{10.1214/10-AOAS401}
\volume{5}
\issue{2A}
\pubyear{2011}
\firstpage{684}
\lastpage{704}

\makeatletter
\newcolumntype{d}[1]{D{.}{.}{#1}}

  \let\sv@tabnotetext\tabnotetext
  \let\sv@tabnotemark@fmt\tabnotemark@fmt
   \long\def\legend#1{{\let\tabnote@indent\leavevmode\sv@tabnotetext[]{}{#1}}}

\makeatother

\begin{document}
\begin{frontmatter}

\title{Point process modeling of wildfire hazard in~Los~Angeles County,
California}
\runtitle{Point process models of wildfires}

\begin{aug}
\author[A]{\fnms{Haiyong} \snm{Xu}\ead[label=e1]{xuhy@ucla.edu}}
and
\author[A]{\fnms{Frederic Paik} \snm{Schoenberg}\corref{}\ead[label=e2]{frederic@stat.ucla.edu}}
\runauthor{H. Xu and F. P. Schoenberg}
\affiliation{University of California, Los Angeles}
\address[A]{Department of Statistics\\
University of California\\
8125 Math-Science Building\\
Los Angeles, California 90095--1554\\
USA\\
\printead{e1}\\
\hphantom{E-mail: }\printead*{e2}} 
\end{aug}

\received{\smonth{4} \syear{2009}}
\revised{\smonth{8} \syear{2010}}

%
\begin{abstract}
The Burning Index (BI) produced daily by the United States government's
National Fire Danger Rating System is commonly used in forecasting the
hazard of wildfire activity in the United States. However, recent
evaluations have shown the BI to be less effective at predicting
wildfires in Los Angeles County, compared to simple point process
models incorporating similar meteorological information. Here, we
explore the forecasting power of a suite of more complex point process
models that use seasonal wildfire trends, daily and lagged weather
variables, and historical spatial burn patterns as covariates, and that
interpolate the records from different weather stations. Results are
compared with models using only the BI. The performance of each model
is compared by Akaike Information Criterion (AIC), as well as by the
power in predicting wildfires in the historical data set and residual
analysis. We find that multiplicative models that directly use weather
variables offer substantial improvement in fit compared to models using
only the BI, and, in particular, models where a distinct spatial
bandwidth parameter is estimated for each weather station appear to
offer substantially improved fit.
\end{abstract}

%

\end{frontmatter}

\section{Introduction}\label{sec1}

This paper explores the use of space--time point process models for the
short-term forecasting of wildfire hazard in Los Angeles County, California.
The region is especially well suited to such an analysis, since the Los
Angeles County Fire Department and Department of Public Works have
collected and compiled detailed records on the locations burned by
large wildfires dating back over a century. The landscape in Los
Angeles County is uniquely vulnerable to high intensity crown-fires,
largely because the predominant local vegetation consists of dense,
highly flammable contiguous chaparral shrub [Keeley (\citeyear{K2000})]. In addition, the dry summers and early autumns in Los
Angeles County are typically followed by high winds known locally as
Santa Ana
winds [Keeley and Fotheringham (\citeyear{KF2003})]. These offshore winds reach
speeds exceeding 100 kph at a relative humidity below 10\%,
and are annual events lasting several days to several weeks, creating the
most severe fire weather in the United States [Schroeder et al. (\citeyear{Setal1964})].

In order to forecast wildfire hazard, the United States National Fire
Danger Rating System (NFDRS), created in 1972, produces several daily
indices that are designed to aid in planning fire control activities on
a fire protection unit [Deeming et al. (\citeyear{DBC1977}); Bradshaw et al. (\citeyear{Betal1983}); Burgan
(\citeyear{B1988})]. These include the Occurrence Index, the Burning Index (BI),
and the Fire Load Index. These indices are derived from three fire
behavior components---a~Spread Component, an Energy Release Component,
and an Ignition Component, that are in turn computed based on fuel age,
environmental parameters (slope, vegetation type, etc.), and
meteorological variables such as wind, temperature, and relative humidity.
Local wildfire management agencies may combine these components in
different ways or calibrate the inherent parameters to adapt the system
to the local environment for wildfire hazard assessment.

Fire managers use this information in making decisions about the
appropriateness of prescribed burning or alerts for increased
preparedness, both in terms of fire suppression staffing and fire
prevention activities. Since fireline intensity is an important factor
in predicting fire containment and the likelihood of fire escape, the
Burning Index is the rating of most interest to many fire managers
[Schoenberg et al. (\citeyear{Setal2008})]. This is especially the case for natural
crown-fire ecosystems such as southern California shrublands, where BI
is commonly employed to assess fire danger [Mees and Chase (\citeyear{MC1991})].
Indeed, in Los Angeles County, as well as at least $90\%$ of counties
nationwide, the BI is the index primarily used by fire department
officials as a measure of overall wildfire hazard, and its use has been
justified largely based on its observed empirical correlation with
wildfire incidence and burn area in different regions [Haines et al.\
(\citeyear{Hetal1983}); Haines, Main and Simard (\citeyear{HMS1986}); Mees and Chase (\citeyear{MC1991});
Andrews, Loftsgaarden and Bradshaw (\citeyear{ALB2003})].
%
However, several recent investigations have shown that the BI is far
from an ideal predictor of wildfire incidence in Los Angeles County;
Schoenberg et al. (\citeyear{Setal2008}) showed that a simple point process model,
which used only the same weather variables as those incorporated by the
BI, vastly outperformed the BI in terms of predictive efficacy in Los
Angeles County, using historical data from 1977--2000. In fact, the
simple model in Schoenberg et al. (\citeyear{Setal2008}) not only offered improvement
in terms of likelihood scores such as the Akaike Information Criterion
(AIC), but the study suggested that substantial improvement in
short-term forecasting could be achieved by the simple model using the
weather variables directly, compared to a point process model that
interpolates BI measurements.

Here, we adopt the same basic modeling framework of Schoenberg et al.
(\citeyear{Setal2008}), but extend the models in two important ways. First, we consider
not only daily weather variables but also additional covariates with
management relevance, such as historical spatial burn patterns and wind
direction, using the directional kernel regression method described in
Schoenberg and Xu (\citeyear{SX2008}). Second, unlike the simple models of Mees and
Chase (\citeyear{MC1991}) and Schoenberg et al. (\citeyear{Setal2008})
that average daily weather variables over weather stations within Los
Angeles County, here we explore models that interpolate the records
from different weather stations, weighting these data based on their
spatial distance from the location where wildfire hazard is to be
estimated. Thus, the models considered here should have more direct
relevance for forecasting wildfire hazard in precise spatial locations
within Los Angeles County, compared to previous work that essentially
averaged weather variables and hazard estimates over Los Angeles County
as a whole. As with Schoenberg et al. (\citeyear{Setal2008}), our results are compared
with models using the BI measurements recorded at each of the weather
stations, so that the effectiveness of the BI in summarizing the
wildfire hazard as a function of the weather variables may be assessed.

While alternative models may be more useful for forecasting long-term
wildfire hazard, that is,\ estimating the number of wildfires occurring
within a month, season, or year, the focus here is on forecasting
short-term wildfire hazard, that is,\ the probability of a wildfire
occurring within a specific day.
To compare the overall performance of the models considered, we employ
diagnostics including likelihood-based numerical summaries such as the
Akaike Information Criterion (AIC), as well as power diagrams
summarizing the predictive efficacy of each model for short-term
forecasting. Residual analysis is also used to highlight specific areas
and times where the performance of a model is poor and to suggest areas
for improvement.

The paper proceeds as follows. Section \ref{sec2} describes the wildfire and
weather data that are used in the analysis. The models used, as well as
methods for their estimation, are outlined in Section \ref{sec3}, and methods
for goodness-of-fit assessment are discussed in Section \ref{sec4}. Section \ref{sec5}
presents the main results, and a discussion is given in Section \ref{sec6}.

\section{Data}\label{sec2}
\subsection{Wildfire data}\label{sec2.1}
Los Angeles County is an ideal test site for models for wildfire
hazard, with detailed wildfire data having been collected and compiled
by various agencies, including the Los Angeles County Fire Department
(LACFD) and the Los Angeles County Department of Public Works, the
Santa Monica Mountains Recreation Area, and the California Department
of Forestry and Fire Protection. Regional records of the occurrence of
wildfires date back to 1878, and include information on each fire,
including its origin date, the polygonal outline of the resulting area
burned, and the centroidal location of this polygon. LACFD officials
have noted that the records prior to 1950 are believed to be complete
for fires greater than $0.405$~km$^2$ (100 acres), and data since 1950
are believed to be complete for fires burning greater than $0.0405$~km$^2$, or 10 acres [Schoenberg et al. (\citeyear{Setal2003a})]. As in Schoenberg et al.
(\citeyear{Setal2008}), our analysis in this paper is focused primarily on models for
the occurrences of the 592 wildfires burning at least 0.0405~km$^2$
recorded between January 1976 and December 2000. The daily burn area is
highly right-skewed and closely follows the tapered Pareto distribution
[Schoenberg, Peng and Woods (\citeyear{SPW2003b})]. For further details, images of the spatial
locations of these wildfires, and information about missing data, see
Peng, Schoenberg and Woods (\citeyear{PSW2005}).

\subsection{Meteorological data}\label{sec2.2}

Since 1976, daily meteorological observations from the Remote Automatic
Weather Stations (RAWS) were archived across the United States. The
analysis here is based on sixteen RAWS located within Los Angeles
County, California. The RAWS record daily measures of many
meteorological variables, including air temperature, relative humidity,
precipitation, wind speed, and wind direction [Warren and Vance
(\citeyear{WV1981})]. Summaries of these records are collected daily at 1300 hr and
transmitted by satellite to a central archiving station. These daily
RAWS data are used as inputs by the NFDRS in order to construct fire
behavior components that are in turn combined to construct the BI. It
should be noted that data were missing on certain days for several of
the 16 RAWS, though the biases resulting from such missing data are
likely to be small; see Peng, Schoenberg and Woods (\citeyear{PSW2005}) for details.

\section{Methodology}\label{sec3}

We follow previous research including Schoenberg et al. (\citeyear{Setal2008}) in
modeling the catalog of wildfire centroids in Los Angeles County as a
realization of a point process that may depend on daily meteorological
variables. We begin with a basic reference model using merely a spatial
background rate and seasonal component, and a model using the Burning
Index in addition to the spatial and seasonal background rates. We then
introduce competing models that use daily meteorological variables
recorded at the RAWS, and extend the research of Schoenberg et al. (\citeyear{Setal2008}) by including additional covariates, such as wind direction and
fuel age. Further, instead of averaging daily weather variables or the
Burning Index over all weather stations within Los Angeles County, here
we explore methods of obtaining an estimated spatial intensity at any
location $x$ on any particular day by interpolating the meteorological
variables from different weather stations, weighting each record based
on its distance from the location $x$ in question.

\subsection{A review of point process modeling}\label{sec3.1}

A spatial-temporal point process $N$ is mathematically defined as a
random measure on a spatial-temporal region $S$, taking values in the
nonnegative integers $\mathbb{Z}^+$ or infinity [Daley and Vere-Jones
(\citeyear{DV2003})]. In this framework the measure $N(A)$ represents the number of
points falling in the subset $A$ of $S$. Since any analytical
spatial-temporal point process is characterized uniquely by its
associated conditional rate (or intensity) $\lambda(s)$, assuming it
exists, modeling of such point processes is typically performed by
specifying a parametric model for this rate. For the case where the
spatial region is planar, for any point $t$ in time and location
$(x,y)$ in the plane, the conditional rate is defined as a limiting
frequency at which events are expected to occur within time range $(t,
t+\Delta t)$ and rectangle $(x, x+\Delta x) \times(y, y+ \Delta y)$,
conditional on the prior history, $H_t$, of the point process up to
time $t$. For references on space--time point processes and conditional
rates, see, for example,\ Daley and Vere-Jones (\citeyear{DV1988}) or Schoenberg, Brillinger and Guttorp (\citeyear{SBG2002}).

Given a parametric function for $\lambda(t, x, y)$, estimates of the
parameters $\theta$ may be obtained by maximizing the log-likelihood
function [see Schoenberg, Brillinger and Guttorp (\citeyear{SBG2002}), page\ 1576, or Daley and
Vere-Jones (\citeyear{DV2003}), page~232, equation~7.24]:
\begin{eqnarray*}\label{eq: MLE}
L(\theta) &=& \int^{T_1}_{T_0} \int_x \int_y \log [ \lambda(t, x, y;
\theta) ] \,dN(t, x, y) - \int^{T_1}_{T_0} \int_x \int_y \lambda(t, x,
y;, \theta) \,dy \,dx \,dt \\
&=& \sum^n_{i=1} \log \lambda(t_i, x_i, y_i; \theta) - \int^{T_1}_{T_0}
\int_x \int_y \lambda(t, x, y; \theta) \,dy \,dx \,dt.
\end{eqnarray*}
In the case of a Poisson process, the intuition behind this formula is that
$\prod _{i=1} ^n \lambda(t_i,x_i,y_i; \theta)$ reflects the likelihood
associated with the observed events, and
$\exp\{\int^{T_1}_{T_0} \int_x \int_y \lambda(t,x,y; \theta) \,dy \,dx \,dt\}
$ represents the probability of no events in any \textit{other} portions
of the spatial-temporal region, the full likelihood is the product of
these two terms, and the logarithm of this product yields $L(\theta)$ above.
Under rather general conditions, the maximum likelihood estimates
(MLEs) are consistent, asymptotically normal, and efficient [Ogata (\citeyear{O1978})], and estimates of their variance can be derived
from the negative of the diagonal elements of the inverse Hessian of
the likelihood function [Ogata (\citeyear{O1978}), Rathbun and Cressie (\citeyear{RC1994})]. In
most cases, explicit solutions for MLEs are not available and iterative
numerical optimization methods are used instead.

\subsection{A simple reference model}\label{sec3.2}

In this analysis we explore several spatial-temporal point process
models for predicting wildfire occurrence rates. As an initial baseline
model, one may consider an inhomogeneous Poisson process, where the
conditional intensity at time $t$ and at location $(x,y)$ depends only
on the season associated with time $t$, as well as the background rate
$m(x,y)$ of wildfires for the location in question. That is, one may
consider a baseline model such as
%
\begin{equation} \label{eqn: model1}
\lambda_1(t, x, y) = \gamma m(x, y) + \alpha S(t),
\end{equation}
where $\gamma$ and $\alpha$ are parameters to be estimated in modeling fitting.

Parametric or nonparametric methods can be used to estimate the
seasonal pattern $S(t)$ and spatial background $m(x,y)$. While
nonparametric methods can be especially flexible for estimating complex
patterns such as spatial burn averages, a possible drawback to such
methods is their potential for overfitting, particularly when the same
data are used for fitting and evaluation of the fit of the model.
As in Schoenberg et al. (\citeyear{Setal2008}), we propose estimating the spatial
background $m(x,y)$ for fires between 1976--2000 by kernel smoothing
the centroidal locations of wildfires recorded during the previous 25
years, that is,\ from January 1950 to December 1975. That is,
\[
m(x, y) = \frac{1}{n_0 \beta_m} \sum _{j=1}^{n_0} K \biggl( \frac{ \| (x, y) -
(x_j, y_j) \| }{\beta_m} \biggr),
\]
where $K$ is a kernel function, $\beta_m$ is a bandwidth to be
estimated in modeling fitting, $(x_j, y_j)$ indicates the spatial
coordinates of the $j$th wildfire between 1950 and 1975, $n_0$ is
the number of observed 1951--1975 wildfire occurrences, and $\| (x, y) -
(x_j, y_j) \|$ is the Euclidian distance between $(x,y)$ and $(x_j,
y_j)$. Standard kernel functions can be used, and attention is usually
limited to functions that are unimodal, symmetric about zero, and that
integrate to~1, such as the Gaussian density of the Epanechnikov kernel
[H\"{a}rdle (\citeyear{H1994})]. It is well known that the results are far more
sensitive to the choice of bandwidth than the choice of kernel
function, and much research has focused on automated methods for
choosing bandwidth parameters, including cross-validation, penalty
functions, and plug-in methods [Silverman (\citeyear{S1986}); H\"{a}rdle (\citeyear{H1994})].
Here, since the data $(x_j, y_j)$ used in the estimation of $m(x,y)$ is
distinct from that used in the rest of the model fitting and in the
evaluation, the problem of overfitting is far less severe, and the
bandwidth parameter may simply be fitted by maximum likelihood.

\begin{figure}

\includegraphics{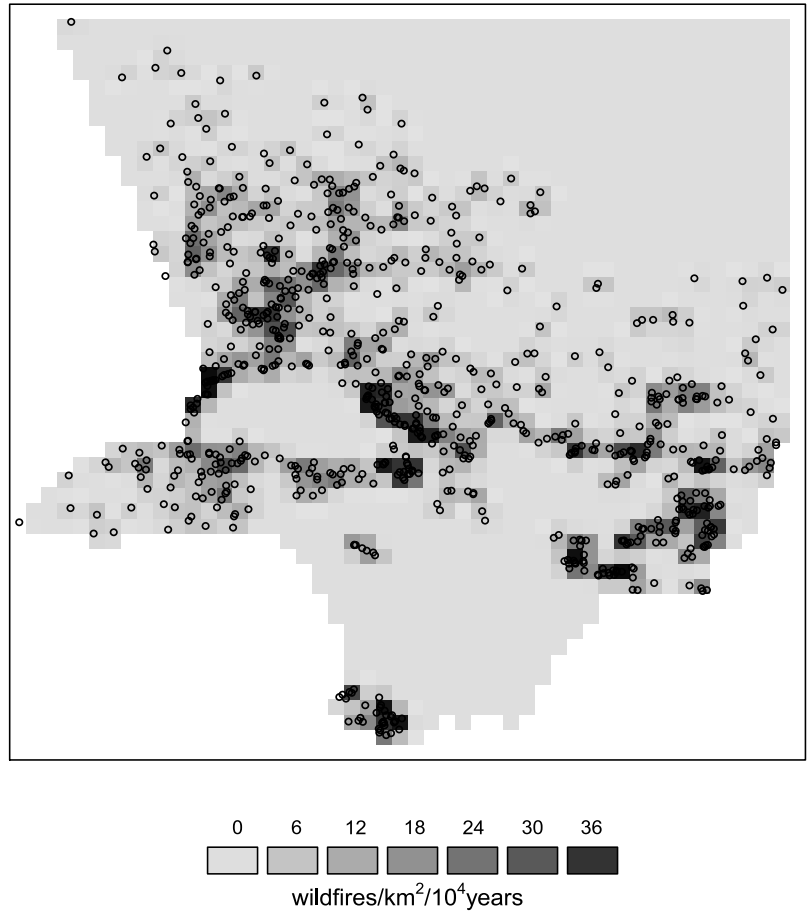}

  \caption{Spatial background rate $m(x,y)$, with centroid locations of wildfires occurring during 1878--1976.
  (The spatial bandwidth $\beta_m$ is 0.6 miles).}\label{fig1}
\end{figure}

Figure \ref{fig1} shows an estimate of the spatial background rate $m(x,y)$,
with bandwidth estimated by maximum likelihood. One sees the general
pattern of fire activity in Los Angeles County during 1951--1975, with
most fires occurring in the Angeles National Forest, as well as parts
of the Los Padres National Forest and the Santa Monica Mountains, while
many other wildfires were located in or near Buckweed, Santa Clarita,
and Glendale, California.

Helmers, Magku and Zitikis (\citeyear{HMZ2003}) propose a kernel-based estimate for the
consistent estimation of a seasonal time series. Here, in order to
safeguard against overfitting, we propose estimating the seasonal
pattern $S(t)$ describing the overall seasonal variation of wildfire
activity in a fashion similar to that used for the spatial background
rate, that is,\ by kernel smoothing the times of wildfires during \textit{previous} years:
\[
S(t) = \frac{1}{n_0 \beta_t} \sum _{j=1}^{n_0} K \biggl( \frac{ T^*(t) -
T^*(t_j) }{\beta_t} \biggr).
\]
In the above equation, $T^*(t)$ represents the date within the year
associated with time $t$, that is,\ $T^*(t)$ is the number of days
since the beginning of the year for time $t$,
$t_j$ is the time of the $j$th wildfire occurrence in the data set
(1950--1975), and $ \beta_t$ is a bandwidth parameter to be estimated.
A~wrapped kernel function $K$ should be used so that, for instance,
January~1 and December 31 are treated as one day apart. The bandwidth
may be estimated by maximum likelihood, fitting the kernel smoothing of
the 1950--1975 data to the 1976--2000 data set. This procedure may be
preferable for relatively small data sets such as the one considered
here in order to prevent overfitting.

\begin{figure}

\includegraphics{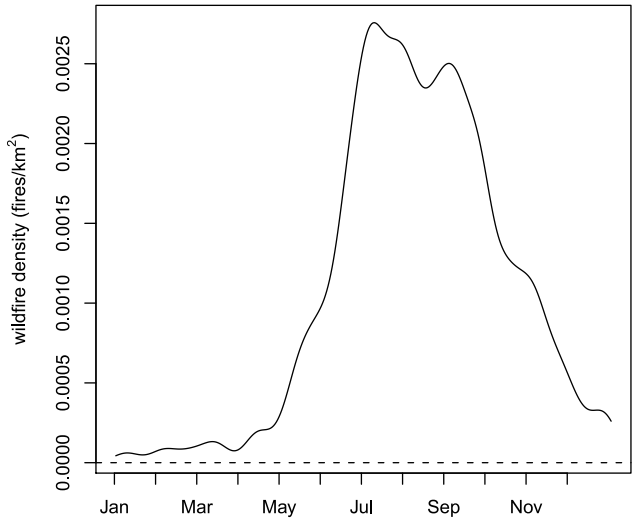}

  \caption{Estimate of the seasonal pattern $S(t)$ for model (\protect\ref{eqn: model1}),
  using kernel regression with a wrapped Gaussian kernel and bandwidth $\hat \beta_t  = 9.86$ days, estimated by MLE.}\label{fig2}
\end{figure}

Figure \ref{fig2} displays the smoothed function $S(t)$ applied to wildfire
incidence in Los Angeles County from January 1950 to December 1975,
with bandwidth estimated by MLE by fitting the resulting function to
wildfire data from 1976--2000. It is evident that the mean number of
wildfires is highest between July and October and rapidly decreases
during November and December, reaching its minimum in January and February.
Schoenberg et al.\ (\citeyear{Setal2008}) pointed out that the Burning Index typically
assumes moderate values in December, January, and February, though few
wildfires occur during these months.

\subsection{A point process model using Burning Index}\label{sec3.3}

To evaluate the potential of the Burning Index (BI) in predicting
wildfire incidence, one may consider a model such as
%
\begin{equation} \label{eqn: model2}
\lambda_2(t, x, y) = \gamma m(x, y) + \alpha S(t) + \mu_{\mathrm{BI}}B(t, x, y)
\end{equation}
for some function $B(t, x, y)$ which interpolates the BI records at
time $t$ and location $(x, y)$, since BI records are only available at
fixed RAWS sites.
Different methods of interpolation are possible. One possibility is to
average the BI records on day $t$, weighing each by the distance
between the RAWS and the location $(x, y)$ in question. That is,
\[
B(t, x, y) = \frac{1}{C_{\mathrm{BI}}} \sum _{s \in S_t} \biggl\{ K \biggl( \frac{ \| (x,
y) - (x_s, y_s) \| }{\beta_{\mathrm{BI}}} \biggr)\operatorname{BI}(t, s) \biggr\},
\]
where $\operatorname{BI}(t,s)$ is the BI value recorded at time $t$ from the $s$th
station, $(x_s, y_s)$ are the coordinates of the $s$th station,
$S_t$ represents the collection of stations for which BI records are
available on day $t$, and $C_{\mathrm{BI}}$ is a normalizing constant given by
\[
C_{\mathrm{BI}} = \sum _{s \in S_t} \biggl\{ K \biggl( \frac{ \| (x, y) - (x_s, y_s) \|
}{\beta_{\mathrm{BI}}} \biggr) \biggr\}.
\]

\subsection{Models using spatial interpolation of meteorological
variables, including wind speed and wind direction}\label{sec3.4}

As an alternative to the model (\ref{eqn: model2}) incorporating BI
measurements, one may instead consider examining the direct impact on
wildfire hazard estimates of meteorological variables used in the
computation of the BI, by replacing the function $B(t, x, y)$ in (\ref
{eqn: model2}) by functions of the meteorological variables themselves.
That is, one may consider models such as
%
\begin{equation} \label{eqn: model3}
\lambda_3 (t, x, y) = \gamma m(x, y) + \alpha S(t) + F_{1}(t, x, y),
\end{equation}
where $F_{1}(t,x,y)$ takes into account the contribution of temperature
($T$), relative humidity ($H$), wind speed ($W$), and precipitation ($P$) at
time $t$~from each RAWS where the data are available.

Since nonlinearities have been detected in the dependence of burn area
on climatic variables [Schoenberg et al.\ (\citeyear{Setal2003a})], one may wish to
avoid simple averaging of the meteorological variables in estimating
wildfire hazard. Instead, one option is to describe the association
between each climatic variable and wildfire burn area by an explicit
function $g$ and weight the information from each RAWS by the distance
to the point $(x, y)$ to be estimated using kernel smoothing. This
suggests a model such as
\begin{eqnarray*}
F_{1} (t, x, y) &=& \frac{\mu_{T}}{C_T} \sum _{s \in S_t} \biggl\{K \biggl( \frac{
\| (x, y) - (x_s, y_s) \| }{\beta_T} \biggr)g_T(T(t, s)) \biggr\} \\
&&{}+ \frac{\mu_{H}}{C_{H}} \sum _{s \in S_t} \biggl\{K \biggl( \frac{ \| (x, y) -
(x_s, y_s) \| }{\beta_{H}} \biggr)g_{H}(H(t, s)) \biggr\} \\
&&{}+ \frac{\mu_W}{C_W} \sum _{s \in S_t} \biggl\{K \biggl( \frac{ \| (x, y) - (x_s,
y_s) \| }{\beta_{W}} \biggr)g_W(W(t, s)) \biggr\} \\
&&{}+ \frac{\mu_{P}}{C_{P}} \sum _{s \in S_t} \biggl\{K \biggl( \frac{ \| (x, y) -
(x_s, y_s) \| }{\beta_{P}} \biggr)g_{P}(P(t, s)) \biggr\},
\end{eqnarray*}
where $T(t, s)$, $H(t, s)$, $W(t, s)$, and $P(t, s)$ are records of
temperature, relative humidity, directed wind speed, and precipitation,
respectively, on day~$t$ at the $s$th RAWS, the parameters $\mu_T,
\mu_{R}, \mu_W$, and $\mu_{P}$ represent weights associated with these
meteorological variables, $\beta_T, \beta_{H}, \beta_W$, and $\beta
_{P}$ are bandwidths to be estimated, and $C_T, C_{H}, C_W$, and
$C_{P}$ are normalizing constants.

Note that the bandwidth parameters are somewhat different here than in
ordinary kernel regression models. While ordinarily in kernel
regression or kernel density estimation bandwidth parameters may not
typically be estimated by maximum likelihood because the likelihood
would tend to increase as the bandwidth shrinks to 0 [Silverman
(\citeyear{S1986})], here this is not the case. Instead, the bandwidth parameters
$\beta_T, \beta_H, \beta_W$, and $\beta_P$ in model (\ref{eqn: model3}) merely control
the spheres of influence of the relative weather stations in terms of
the impact of each on wildfire hazard. That is, if $\beta_T$ is small,
for instance, then each RAWS station's recorded temperature will affect
the wildfire incidence more locally, whereas if $\beta_T$ is very
large, then the wildfire hazard at any particular location will depend
more closely on the average temperature throughout Los Angeles County.

\begin{figure}

\includegraphics{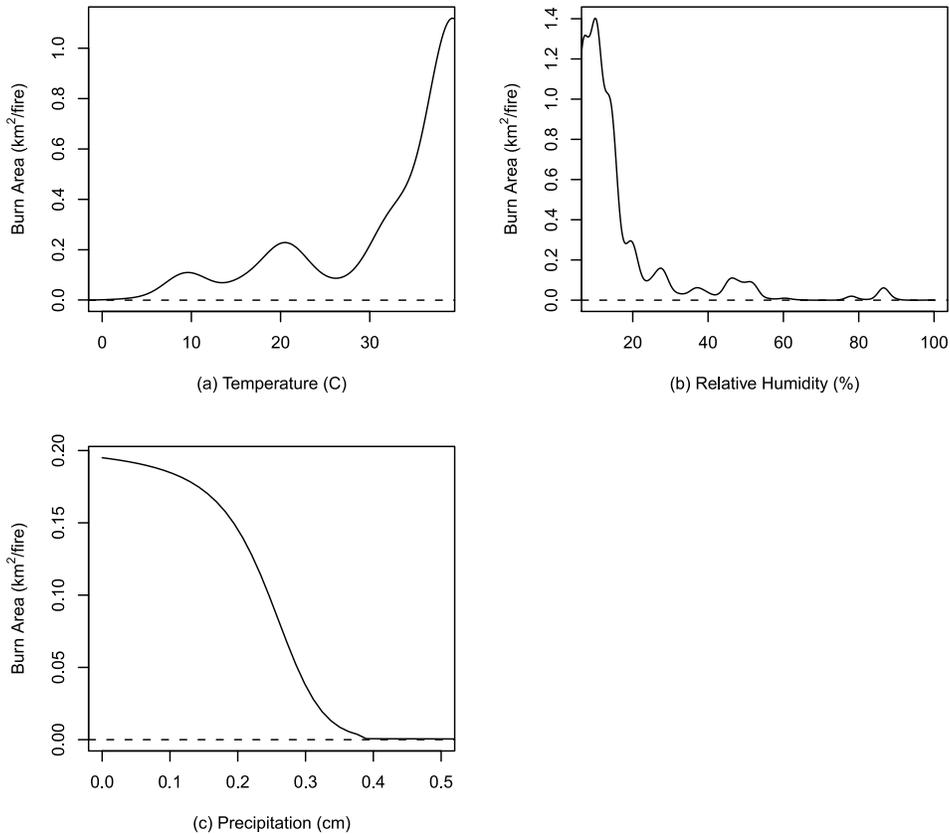}

  \caption{Kernel regression estimates of the
  relationship between daily burn area and \textup{(a)}~temperature, \textup{(b)}
  relative humidity, and \textup{(c)} precipitation. Gaussian kernels are used,
  bandwidths are estimated by cross-validation, and edge correction is
  performed via reflection [Silverman (\protect\citeyear{S1986})].}\label{fig3}
\end{figure}

Functional forms can be suggested for $g_T, g_{H}, g_{W}$, and $g_{P}$,
by individually examining the empirical relationship between daily area
burned and each of these variables. In order to smooth these
relationships, one possibility would be to use local linear regression
or segmented regression, since the relationships between wildfire burn
area and temperature, precipitation, and other weather variables appear
to have thresholds [Schoenberg et al. (\citeyear{Setal2003a})]. Another possibility is
to use kernel regression of daily area burned on the average
temperature, relative humidity, and precipitation over all RAWS,
respectively. For instance, the impact of temperature may be estimated via
\[
g_T(T) =\frac{ \sum _{j=1}^{n_1} \{ K ( |T-T_j|/{h_T} )A_j \}
}{\sum _j \{ K (|T-T_j|/{h_T} ) \}},
\]
where $A_j$ is the area burned on the $j$th day during 1976--2000,
$T_j$ is the average temperature readings over all RAWS on that day,
$h_T$ is the bandwidth of the kernel regression which can be selected
by methods such as cross-validation or the plug-in method [Silverman
(\citeyear{S1986})], and $n_1$ is the number of days with records during this
period. Figure \ref{fig3} displays such kernel regression estimates of $g_T$,
$g_{H}$, and $g_{P}$. Not surprisingly, one sees that daily area burned
generally increases as temperature increases, and decreases as relative
humidity and precipitation increase, though some local fluctuations are
seen in the kernel regressions on temperature and relative humidity.
These fluctuations are likely attributable to the high variability of
the estimates due to the relatively small sample of large fires
contained in the catalog.

Special care should be taken in estimating $g_W$, since wind is
directional, and this direction may provide important information
related to wildfire incidence.
One possible way to estimate the relationship between daily area burned
and directional wind speed is via directional kernel regression, as
outlined in Schoenberg and Xu (\citeyear{SX2008}).
An example of a two-dimensional directional kernel is the von Mises
distribution suggested
by Mardia and Jupp (\citeyear{MJ2000}):
\[
vM(\theta; \mu, \kappa) = \frac{1}{2\pi I_0(\kappa)} e^{\kappa\cos
(\theta- \mu)},
\]
where $I_0$ denotes the modified Bessel function of the first kind and
order 0, $\mu$ is the directional center, and $\kappa$ is known as the
concentration parameter.
Following Schoenberg and Xu (\citeyear{SX2008}), the corresponding two-dimensional
kernel regression function
$g_W$ would then be estimated via
\[
g_W(W, \theta) = \frac{ \sum _{j=1}^{n_1} \{ K ( { |W - W_j|}/
{h_{W}} ) vM ( \theta- \theta_j; \mu_0, \kappa_0 ) A_j \}} { \sum
_{j=1}^{n_1} \{ K ( { |W- W_j|} /{h_{W}} ) vM ( \theta- \theta_j;
\mu_0, \kappa_0 ) \}},
\]
where $W_j$ and $\theta_j$ represent the mean wind speed and wind
direction, respectively, on day $j$.
Cross-validation can be used to optimize the estimates of~$h_{sp}$, $\mu
_0$, and~$\kappa_0$.

\begin{figure}

\includegraphics{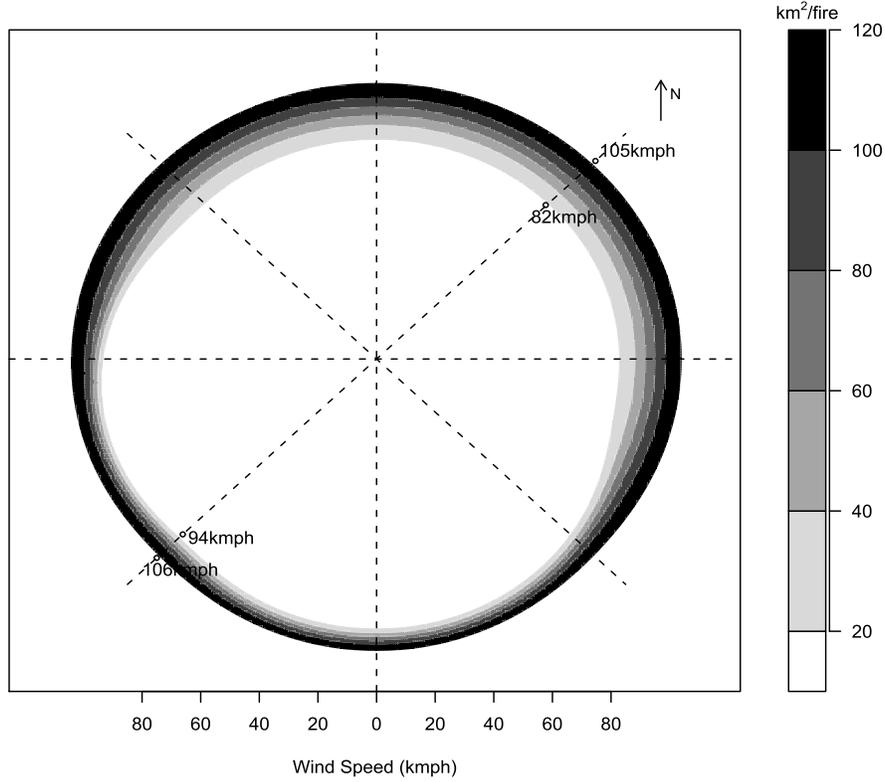}

  \caption{Two-dimensional kernel regression of burn
  area versus wind speed and wind direction. Wind speed and wind
  direction are represented in a polar system: wind speed is
  represented by the distance to the center and wind direction is
  represented by the angle. The grey scale represents the smoothed
  average wildfire burn area during 1976--2000.}\label{fig4}
\end{figure}

Figure \ref{fig4} displays a kernel regression estimate of the relationship
between daily burning area and daily mean wind direction, weighted by
wind speed, averaged over all $16$ RAWS stations.
The sharp increase in mean wildfire burn area, indicated by darker
shading in Figure \ref{fig4}, is very strongly associated with higher wind
speeds. In addition, one sees from Figure \ref{fig4} the extent to which winds
from the northeast, which are often warm, dry Santa Ana winds, are
associated with higher burn areas.
Since the impact on average wildfire burn area of wind direction might
be different at distinct weather stations, one might wish to estimate
$16$ distinct kernel regression functions~$g_W^{(s)}$, one for each
RAWS station $s$.

The model (\ref{eqn: model3}) described above is additive in each of
the weather variables, implying that an extreme value in only one
weather variable may lead to a high estimate of wildfire hazard on the
corresponding day, which might be questionable. For instance, one would
expect few large wildfires occur on days when temperatures are
extremely high yet there is some moderate amount of precipitation and
relative humidity. An alternative approach is to use a multiplicative
component instead, where once again
%
\begin{equation} \label{eqn: model4}
\lambda_4 (t, x, y) = \gamma m(x, y) + \alpha S(t) + \mu F_{2}(t, x, y),
\end{equation}
where now the fire weather $(F)$ term has the multiplicative form
\[
F_{2} (t, x, y) = \frac{1}{C_2} \sum _{s \in S_t}\!\biggl\{\! K \biggl( \frac{ \| (x,
y) - (x_s, y_s) \| }{\beta_{2}} \biggr)g_T(t, s)g_{H}(t, s) g_W(t, s)g_{P}(t,
s) \!\biggr\}.
\]

\subsection{Models allowing records at different RAWS to have different
relationships with wildfire hazard}\label{sec3.5}

In both model (\ref{eqn: model3}) and model (\ref{eqn: model4}), $g_T,
g_{H}$, and $g_{P}$ may be estimated using kernel regression of burn
area on average temperature, relative humidity, and precipitation over
all weather stations, where each station has the same regression
function. However, because of differences in the locations of the
weather stations, including differing altitudes of these stations, some
stations may have lower average temperatures or higher relative
humidities than others throughout the year. Hence, a particular
temperature and relative humidity at one station might indicate a~very
different wildfire hazard than the same values observed at a different
RAWS station. In order to deal with this, one may estimate distinct
kernel regression curves for each RAWS in model (\ref{eqn: model3}) and model (\ref{eqn: model4}). That
is, one might consider
%
\begin{equation} \label{eqn: model5}
\lambda_5 (t, x, y) = \gamma m(x, y) + \alpha S(t) + F _{3}(t, x, y),
\end{equation}
where
\begin{eqnarray*}
F_{3} (t, x, y) &= &\frac{\mu_T}{C_T} \sum _{s \in S_t} \biggl\{ K \biggl( \frac{ \|
(x, y) - (x_s, y_s) \| }{\beta_T} \biggr)g^{(s)}_T(t, s) \biggr\} \\
&&{}+ \frac{\mu_{H}}{C_{H}} \sum _{s \in S_t} \biggl\{ K \biggl( \frac{ \| (x, y) -
(x_s, y_s) \| }{\beta_{H}} \biggr)g^{(s)}_{H}(t, s) \biggr\} \\
&&{}+ \frac{\mu_W}{C_W} \sum _{s \in S_t} \biggl\{K \biggl( \frac{ \| (x, y) - (x_s,
y_s) \| }{\beta_{W}} \biggr)g^{(s)}_W(t, s) \biggr\} \\
&&{}+ \frac{\mu_{P}}{C_{P}} \sum _{s \in S_t} \biggl\{K \biggl( \frac{ \| (x, y) -
(x_s, y_s) \| }{\beta_{P}} \biggr)g^{(s)}_{P}(t, s) \biggr\}
\end{eqnarray*}
or
%
\begin{equation}\label{eqn: model6}
\lambda_6 (t, x, y) = \gamma m(x, y) + \alpha s(t) + \mu F_{4}(t, x, y),
\end{equation}
where
\begin{eqnarray*}
&&F_{4} (t, x, y)\\
&&\qquad = \frac{1}{C} \sum _{s \in S_t} \biggl\{ K \biggl( \frac{ \| (x,
y) - (x_s, y_s) \| }{\beta} \biggr)g^{(s)}_T(t, s)g^{(s)}_{H}(t, s)
g^{(s)}_W(t, s)g^{(s)}_{P}(t, s) \biggr\}.
\end{eqnarray*}
The kernel regression functions such as $g^{(s)}_T(t,s)$ for each
station $s$ may be estimated as in models (\ref{eqn: model3}) and (\ref{eqn: model4}), that is,\ by
kernel regression of the total daily burn area in Los Angeles County
against the temperature at station~$s$.

\subsection{Incorporating fuel age}\label{sec3.6}

One may further improve the models by adding fuel age as a covariate.
Fuel age, or its proxy, the time since the location's last recorded
burn, appears to have a nonlinear, threshold-type relationship with
burn area [Peng and Schoenberg (\citeyear{PS2008})]. Indeed, burn area appears to
increase steadily with fuel age up to ages of approximately 20--30 years
[Peng and Schoenberg (\citeyear{PS2008})]. This suggests incorporating the
contribution of fuel age into model (\ref{eqn: model5}) by a truncated linear function,
that~is,
%
\begin{equation} \label{eq7}
\lambda_7 (t, x, y) = \gamma m(x, y) + \alpha s(t) + F_{3} (t, x, y) +
\mu_{D} \min\{D (t, x, y), \psi\},
\end{equation}
where $D(t, x, y)$ is the fuel age at the space--time pixel $(t, x, y)$,
and where~$\psi$ is an upper truncation time.
Fuel age may be incorporated similarly into model (\ref{eqn: model6}) as well:
%
\begin{equation}\label{eq8}
\qquad \lambda_8 (t, x, y) = \gamma m(x, y) + \alpha S(t) + \mu F _{4}(t, x,
y)+\mu_{D}\min\{D(t, x, y), \psi\}.
\end{equation}

\section{Model assessment}\label{sec4}

Equations (\ref{eqn: model1})--(\ref{eq8}) describe eight point process models that may be used to
predict wildfire hazard at any time and location within Los Angeles
County. In order to compare the performance of these models, one
commonly used method is the Akaike Information Criterion (AIC), which
is defined as $-2L(\beta)+2p$, where $L(\beta)$ is the log-likelihood
and $p$ is the number of fitted parameters in the model. Smaller values
of AIC indicate better fit. The AIC makes a good trade-off between
model complexity and overfitting by rewarding a higher likelihood while
penalizing the addition of more parameters [Akaike (\citeyear{A1977})].

The predictive capacity of competing point process models may also be
compared by examining the models' performance on the 1976--2000 wildfire
data, as suggested in Schoenberg et al.\ (\citeyear{Setal2008}).
Consider a grid of space--time cells, 
with each cell's center separated by some distance $\Delta d$ in space
and a temporal distance $\Delta_t$, and let these cells represent
locations and times where alarms may potentially be issued.
For any such space--time point $(t, x, y)$, one may compute the
estimated conditional intensity $\hat{\lambda}(t, x, y)$ for a
particular model. Consider issuing an alarm if the value of $\hat
{\lambda}(t, x, y)$ is above some certain threshold. We say the alarm
is successful if a wildfire occurs within the cell; otherwise, it is a
false alarm.
The false positive rate of the alarms, defined as the proportion of
cells without wildfires where $\hat\lambda$ exceeded the alarm
threshold, can be compared to the
true positive rate, that is,\ the proportion of wildfires occurring in
cells where $\hat\lambda$ exceeded the alarm threshold, using
traditional Receiver Operating Characteristic (ROC) curves. Each
possible alarm threshold represents a single point on the ROC curve,
and the resulting curve summarizes the potential efficacy of a model in
forecasting wildfires.

While numerical likelihood scores such as AIC and ROC curves can be
useful in evaluating the overall performance of a point process model,
neither method is useful at identifying particular times and locations
where a model fits poorly or suggesting ways in which a model might be improved.
For these purposes, it is useful to inspect plots of residuals, which
may be defined as the difference between the number of events occurring
in a certain space--time interval and the integral of the estimated
conditional intensity over the same interval [Baddeley et al. (\citeyear{Betal2005})].
Negative residuals indicate overestimates of wildfire hazard, and very
large residuals indicate places and times where the model
underestimated wildfire hazard.

\section{Results}\label{sec5}

The maximum likelihood estimates of the parameters for the models (\ref{eqn: model1})--(\ref{eq8}) are listed in Tables \ref{table1} and \ref{table2}.
In Tables \ref{table1} and \ref{table2}, the parameter~$\psi$ was fixed at 22 years for
models 7 and 8, based on Peng and Schoenberg (\citeyear{PS2008}); this parameter was
also fit by maximum likelihood, yielding very similar results, so, for
simplicity, here we report the fit of the model with~$\psi$ fixed at 22 years.
The bandwidths in spatial background $\beta_m$ range from 0.25~km to
1.20~km and the bandwidth in the seasonal component fall within 8.6 to
34.1 days. 
The bandwidths related to spatially kernel smoothing the weather
variables range from 0.024~km to 0.40~km in models (\ref{eqn: model3}), (\ref{eqn: model5}), and~(\ref{eq7}),
with the smallest value for wind speed in model (\ref{eq7}) and the largest
value corresponding to relative humidity in model (\ref{eqn: model3}).
As mentioned in Section \ref{sec3}, these bandwidths can perhaps be interpreted
as reflecting the scales of influence of the weather variables in terms
of their effect on wildfire incidence.\looseness=-1

\begin{table}%
\tabcolsep=0pt
\caption{Maximum likelihood estimates of scaling parameters}
\label{table1}%
\vspace*{-7pt}
\begin{tabular*}{\tablewidth}{@{\extracolsep{\fill}}ld{2.3}d{2.4}d{2.4}d{2.4}d{2.4}d{2.4}d{2.4}cd{2.5}@{\hspace*{-2pt}}}
\hline
\textbf{Model} & \multicolumn{1}{c}{$\bolds{\gamma}$} & \multicolumn{1}{c}{$\bolds{\alpha}$} &
\multicolumn{1}{c}{$\bolds{\mu_{B}}$} & \multicolumn{1}{c}{$\bolds{\mu_T}$} & \multicolumn{1}{c}{$\bolds{\mu_{H}}$} &
\multicolumn{1}{c}{$\bolds{\mu_W}$} & \multicolumn{1}{c}{$\bolds{\mu_{P}}$} & \multicolumn{1}{c}{$\bolds{\mu}$} & \multicolumn{1}{c@{\hspace*{-2pt}}}{$\bolds{\mu_{D}}$}\\
\hline
(\ref{eqn: model1}) & 24 & 2.0 & & & & & & &\\
& (1.3) & (0.20) & & & & & &\\
(\ref{eqn: model2}) & 4.9 & 0.65 & \multicolumn{1}{c}{$8.6\times10^{-4}$} & & & & & &\\
& (0.26) & (0.039) & \multicolumn{1}{c}{($2.0\times10^{-5}$)} & & & & & \\
(\ref{eqn: model3}) & 6.4 & 0.66 & & 0.18 & 0.21 & 1.0 & 0.15 & &\\
& (0.64) & (0.043) & & (0.017) & (0.012) & (0.071) & (0.010) & & \\
(\ref{eqn: model4}) & 6.4 & 2.7 & & & & & & $2.1\times10^3$&\\
& (0.31) & (0.24) & & & & & & (130) & \\
(\ref{eqn: model5}) & 13 & & 0.60 & 0.58 & 0.19 & 17 & 2.1 & &\\
& (0.84) & & (0.051) & (0.054) & (0.011) & (0.85) & (0.14) & & \\
(\ref{eqn: model6}) & 12 & 1.0 & & & & & & $5.1\times10^3$ &\\
& (0.15) & (0.057) & & & & & & (260) & \\
(\ref{eq7}) & 6.9 & 0.60 & & 0.19 & 0.21 & 52 & 0.71 & & 1.0 \\
& (0.44) & (0.054) & & (0.018) & (0.020) & (2.4) & (0.068) & & (0.066)\\
(\ref{eq8}) & 4.8 & 0.54 & & & & & & $1980\times10^3$ & 0.10\\
& (0.37) & (0.049) & & & & & & ($85\times10^3$) & (0.0094) \\
\hline
\end{tabular*}
\legend{All entries have been multiplied by $10^3$ for brevity.}
\vspace*{-7pt}
\end{table}

\begin{table}[b]
\vspace*{-7pt}
\tabcolsep=0pt
\caption{Maximum likelihood estimates of bandwidth parameters}\label{table2}%
\vspace*{-7pt}
\begin{tabular*}{\tablewidth}{@{\extracolsep{\fill}}ld{2.6}d{2.3}d{2.6}d{2.6}d{2.5}d{2.4}d{2.6}d{2.6}@{\hspace*{-2pt}}}
\hline
\textbf{Model} & \multicolumn{1}{c}{$\bolds{\beta_m}$ \textbf{(km)}} & \multicolumn{1}{c}{$\bolds{\beta_t}$ \textbf{(day)}} &
\multicolumn{1}{c}{$\bolds{\beta_{B}}$ \textbf{(km)}} & \multicolumn{1}{c}{$\bolds{\beta_T}$ \textbf{(km)}} &
\multicolumn{1}{c}{$\bolds{\beta_{H}}$ \textbf{(km)}} & \multicolumn{1}{c}{$\bolds{\beta_W}$ \textbf{(km)}} &
\multicolumn{1}{c}{$\bolds{\beta_{P}}$ \textbf{(km)}} & \multicolumn{1}{c@{\hspace*{-2pt}}}{$\bolds{\beta}$ \textbf{(km)}} \\
\hline
(\ref{eqn: model1}) & 1.20 & 9.86 & & & & & &\\
& (0.004) & (2.8) & & & & & & \\
(\ref{eqn: model2}) & 0.92 & 8.64 & 0.40 & & & & &\\
& (0.00077) & (3.30) & (0.0055) & & & & & \\
(\ref{eqn: model3}) & 0.36 & 29.6 & & 0.37 & 0.40 & 0.30 & 0.24 &\\
& (0.002) & (5.6) & & (0.002) & (0.001) & (0.004) & (0.001) & \\
(\ref{eqn: model4}) & 0.92 & 8.6 & & & & & & 0.03 \\
& (0.1) & (2.3) & & & & & & (0.003) \\
(\ref{eqn: model5}) & 0.34 & 34 & & 0.31 & 0.20 & 0.04 & 0.28 &\\
& (0.002) & (8.3) & & (0.001) & (0.001) & (0.002) & (0.0005) & \\
(\ref{eqn: model6}) & 0.25 & 20 & & & & & & 0.19 \\
& (0.025) & (9.7) & & & & & & (0.003) \\
(\ref{eq7}) & 0.46 & 19 & & 0.36 & 0.39 & 0.024 & 0.20 &\\
& (0.00069) & (5.2) & & (0.00071) & (0.0011) & \multicolumn{1}{c}{(3.2e$-$8)} & (0.00025) &\\
(\ref{eq8}) & 0.99 & 13 & & & & & & 0.037\\
& (0.17) & (2.6) & & & & & & (0.00064) \\
\hline
\end{tabular*}
\end{table}

Table\ \ref{table3} presents the relative AIC values for models (\ref{eqn: model1})--(\ref{eq8}). For
simplicity and ease of presentation, the AIC for the best fitting model
(\ref{eq8}) has been subtracted from the AIC of each model. It is evident that
the BI model (\ref{eqn: model2}) offers very substantial improvement over the baseline
model (\ref{eqn: model1}). However, all the other models that use weather information
directly have much better fits than the BI model (\ref{eqn: model2}). The
multiplicative model (\ref{eq8}) with fuel age appears to offer by far the best
fit among these models, without using the BI directly, and only
involves one more fitted parameter than the BI model.

\begin{figure}[b]
\vspace*{-3pt}
\includegraphics{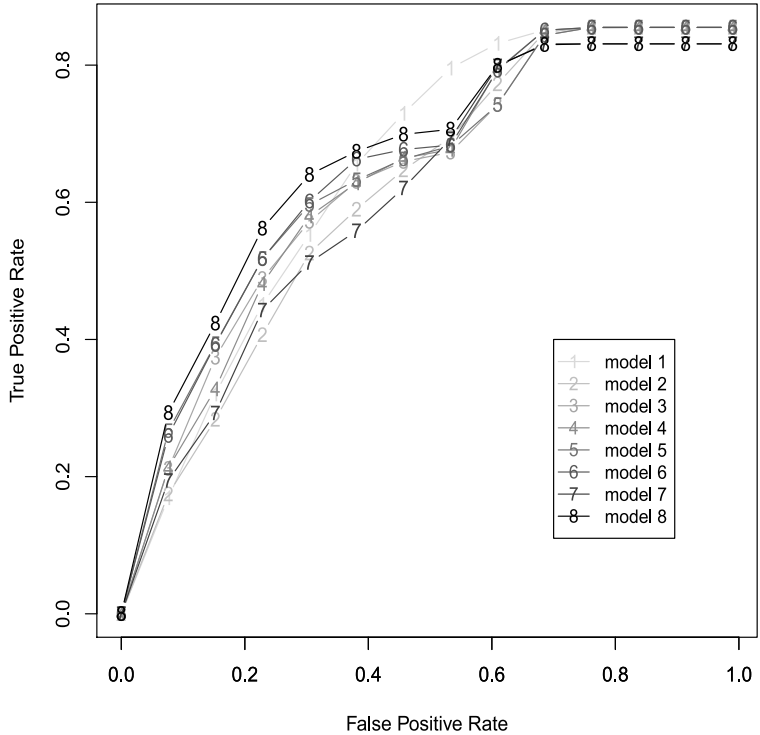}
\vspace*{-5pt}
  \caption{ROC curves for models (\protect\ref{eqn: model1})--(\protect\ref{eq8}),
with $\Delta_t = 1.0$ day and $\Delta d = 4.0$ km.}\label{fig5}
\end{figure}

Figure \ref{fig5} shows a comparison of the predictive efficacy of models
described in Section \ref{sec3}. Models (\ref{eqn: model6}) and (\ref{eq8}) vastly outperform the other
models. The performance was evaluated using a regular space--time grid,
so that each alarm's success or failure was evaluated over a space--time
window with $\Delta d = 4.0$~km and $\Delta t = 1.0$ day.

\begin{table}
\caption{Relative AIC values}
\label{table3}%
\vspace*{-5pt}
\begin{tabular}{@{}lcccccccc@{}}
\hline
Model & (\ref{eqn: model1}) & (\ref{eqn: model2}) & (\ref{eqn: model3}) & (\ref{eqn: model4}) & (\ref{eqn: model5}) & (\ref{eqn: model6}) &(\ref{eq7}) & (\ref{eq8})\\
Relative AIC & 3783 & 2859 & 2509 & 2631 & 2223 & 2209 & 1308 & 0\\
p & 4 & 6 & 12 & 6 & 12 & 6 & 13 & 7\\
\hline
\end{tabular}
\vspace*{-4pt}
\end{table}

For any given success rate, the models that directly use the
meteorological data offer substantially fewer false alarms than the
model (\ref{eqn: model2}) that uses the BI.
For instance, for a false positive rate fixed at $0.08$, model (\ref{eq8})
correctly signals approximately $29\%$ of the wildfires in the data
set, compared to $18\%$ for model (\ref{eqn: model2}). Model (\ref{eqn: model6}), which uses only
temperature, relative humidity, wind speed, wind direction, and
precipitation, but does not use fuel age, signals nearly $25\%$ of the
wildfires correctly with a false positive rate of $0.08$.
Note that this method of evaluating predictive efficacy over a fine
grid of spatial-temporal locations is rather cumbersome for models
(\ref{eq7})--(\ref{eq8}), due to the need to individually estimate the fuel age
associated with each wildfire, with respect to each spatial-temporal
grid location and time, and each such evaluation requires a rather
burdensome computation described in Peng and Schoenberg (\citeyear{PS2008}).

\begin{figure}

\includegraphics{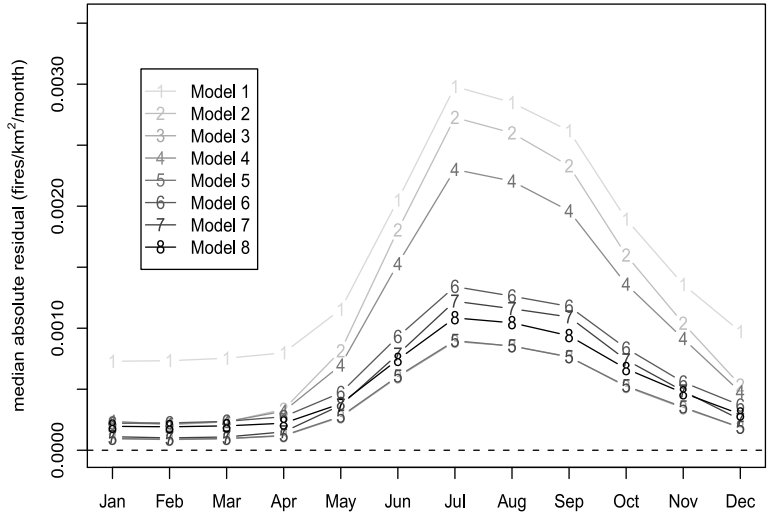}
\vspace*{-5pt}
  \caption{Median absolute value of residuals, by month, using a
space--time grid of $25.6\mbox{ sqkm} \times 30.0$~days.}\label{fig6}
\vspace*{-5pt}
\end{figure}

\begin{figure}

\includegraphics{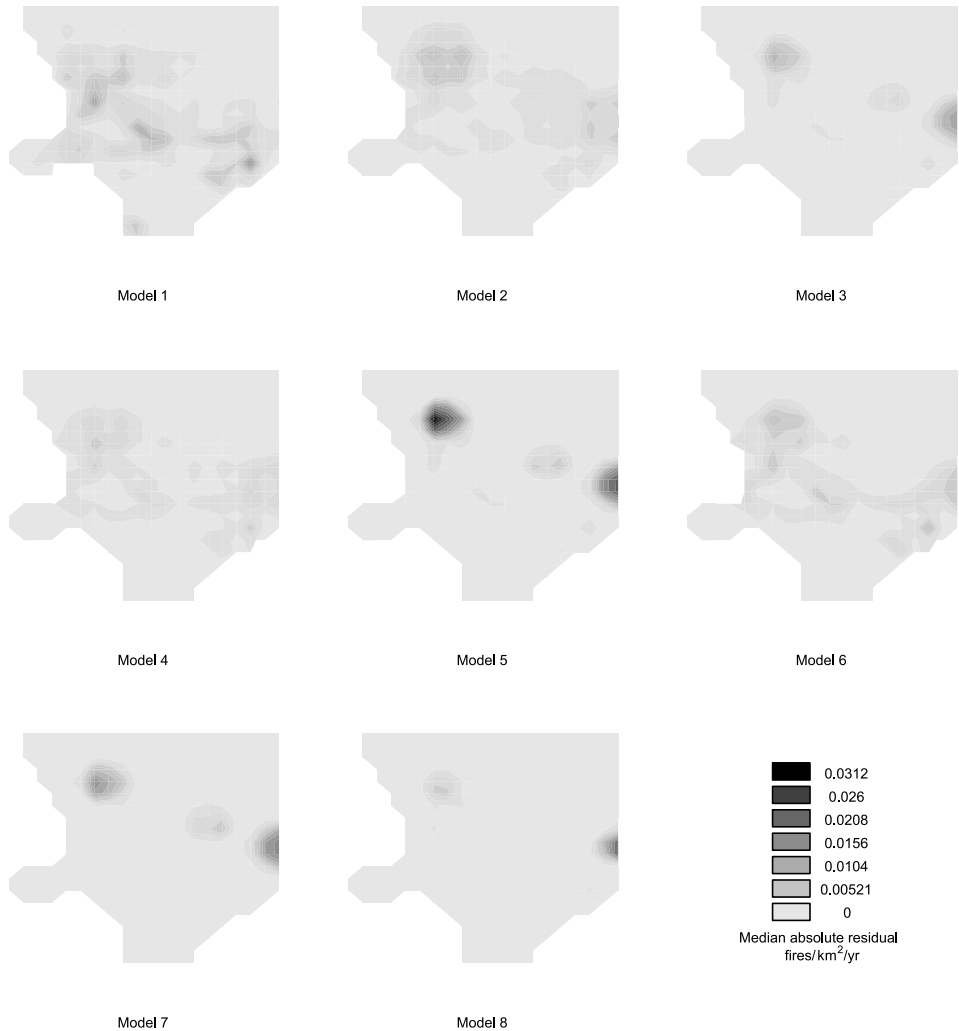}

  \caption{Median absolute value of residuals, by location, using a
space--time grid of $25.6\mbox{ sqkm} \times 30.0$~days.}\label{fig7}
\end{figure}

The fit of the models can be evaluated by examining their
spatial-temporal residuals over a relatively coarse grid. For instance,
Figure \ref{fig6} shows the medians of the absolute values of the residuals in
each month, for models (\ref{eqn: model1})--(\ref{eq8}), where each residual is computed over a
space--time grid of $25.6$ sqkm $\times$ $30.0$ days.
It is evident that models (\ref{eqn: model3}), (\ref{eqn: model5}), (\ref{eqn: model6}), (\ref{eq7}), and (\ref{eq8}) outperform the
other three models, especially in the late Summer and Fall months when
most wildfires in Los Angeles County occur. The months of October and
November are especially critical, since Santa Ana winds prevail and can
cause catastrophic wildfires.
Figure \ref{fig7} shows a spatial plot of the medians of the absolute values of
the residuals, over all months.
From Figure \ref{fig7} one sees that the eight models perform surprisingly
comparably in terms of the median absolute residual, though models
(\ref{eqn: model3})--(\ref{eq8}), which use meteorological data directly, have better
performance than model (\ref{eqn: model2}) in the sense that the corresponding
residuals are generally closer to zero in most areas. These residual
plots also indicate that several of the models may require improvement
in the Northwest portion of the Angeles National Forest, as well as
near the border with San Bernardino County on the Eastern side of Los
Angeles County. From the residuals in Figure \ref{fig7}, one can observe that
the residuals for models (\ref{eqn: model5}), (\ref{eq7}), and (\ref{eq8}) are highly concentrated
around zero except a few large values that occur in the Northwest part
and east part of Los Angeles County.


\section{Discussion}\label{sec6}

The models explored here use the identical information recorded at the
RAWS stations and used as inputs into the computation of the BI. Hence,
it seems relevant to compare the fit of such models with comparable
spatial interpolations of BI measurements, and the fact that the models
using the weather variables directly appear to offer superior fit
suggests that the BI may not be effective as a short-term forecasting
measure of wildfire hazard in Los Angeles County.

It should be noted that the empirical relationship between a fire
danger rating index such as the BI and wildfire incidence is only one
way to evaluate the effectiveness of such an index; alternatives may
include assessing the cost-effectiveness of staffing or other decisions
made based on the index. Furthermore,
the use of fire danger ratings by fire department officials for
wildfire suppression and prevention activities may confound the
empirical relationship between fire danger ratings and observed
wildfire activity. Nevertheless, most evaluation studies of fire danger
rating systems relate such indices to ultimate fire responses,
including fire incidence and fire size. Indeed, Andrews and Bradshaw
(\citeyear{AB1997}), whose work was instrumental in the current implementation of
the BI, suggested that the value of a fire danger index be evaluated
according to its relationship with fire activity, which may be defined
as the incidence of large wildfires. Such empirical relationships have
been used as support for the use of such rating systems for predictive
purposes [Haines et al. (\citeyear{Hetal1983}); Haines, Main and Simard (\citeyear{HMS1986}); Mees and Chase
(\citeyear{MC1991}); Mandallaz and Ye (\citeyear{MY1997a}, \citeyear{MY1997b});
Viegas et al. (\citeyear{Vetal1999}); Andrews, Loftsgaarden and Bradshaw (\citeyear{ALB2003})]. The results here suggest that, for the purpose of
forecasting wildfire hazard, point process models using RAWS records
and previous wildfire activity as covariates may represent a~promising
alternative to existing indices that use essentially the same information.

However, we must emphasize that the point process models proposed here
remain rather simplistic and could potentially be improved by
incorporating a host of other important variables, such as detailed
vegetation type, vegetation cover, soil characteristics, other weather
variables such as cloud cover and lightning, as well as human factors
such as land use and public policy. The exclusion of such variables
from this analysis is solely motivated by our aim to optimize forecasts
given the same remote, automatically-recorded information used in the
computation of the BI. The models considered here could also perhaps be
improved in various ways. For instance, one might allow long-term
temporal trends and/or allow the seasonal component to vary from year
to year. In addition, one may consider
estimating the kernel function in models (\ref{eqn: model5}) and (\ref{eqn: model6}) for each station
using only local wildfires close to the corresponding station, or
perhaps by some more sophisticated weighting scheme where nearby fires
are given higher weight in the estimation of this function. Because
daily burn areas are right-skewed [Schoenberg, Peng and Woods (\citeyear{SPW2003b})], perhaps
kernel regressions where the response variable is some transformation
of the daily burn area might yield superior results.
An additional important direction for future work is the exploration of
similar point process models for wildfire occurrences in other
locations and for other vegetation types or alternative wildfire
regimes, as well as the use of such models for actual \textit{prospective}
predictions of wildfire activity, rather than merely the empirical
assessment of goodness of fit to historical data.

\section*{Acknowledgments}

Thanks to Larry Bradshaw at the USDA Forest Service for generously
providing us with RAWS data and helping us to process it. Thanks also
to James Woods, Roger Peng, and members of the LACFD and LADPW
(especially Mike Takeshita, Frank Vidales, and Herb Spitzer) for
sharing their data and expertise.


\printaddresses

\end{document}